# Self-adaptive Privacy Concern Detection for User-generated Content

Xuan-Son Vu, Lili Jiang

Department of Computing Science, Umeå University
`sonvx@cs.umu.se, lili.jiang@cs.umu.se`

**Abstract.** To protect user privacy in data analysis, a state-of-the-art strategy is differential privacy in which scientific noise is injected into the real analysis output. The noise masks individual's sensitive information contained in the dataset. However, determining the amount of noise is a key challenge, since too much noise will destroy data utility while too little noise will increase privacy risk. Though previous research works have designed some mechanisms to protect data privacy in different scenarios, most of the existing studies assume uniform privacy concerns for all individuals. Consequently, putting an equal amount of noise to all individuals leads to insufficient privacy protection for some users, while over-protecting others. To address this issue, we propose a self-adaptive approach for privacy concern detection based on user personality. Our experimental studies demonstrate the effectiveness to address a suitable personalized privacy protection for cold-start users (i.e., without their privacy-concern information in training data).

## 1 Introduction

Recent advances in artificial intelligence (AI) have opened many new possibilities (e.g., data analytics, autonomous systems), however, they pose the risk regarding privacy and security [1]. Together with the advances in AI and the increasing volume of user-generated content (UGC) such as social network data and medical data, privacy concern on personal data becomes even more critical [2]. Especially, the cross-disciplinary studies have been conducted with the need of integrating personal data from multiple sources (e.g., studies by combining diet data, work condition data, and health data). This data integration dramatically increases the risk of privacy leakage. For example, Narayanan et al. [3] de-anonymized the published Netflix Prize data by matching with IMDB data[1]. Moreover, not only connecting to external sources will reveal privacy but also internal model parameters. Fredrikson et al. [4] used hill-climbing algorithm on the output probabilities of a computer-vision classifier to reveal individual faces from the training data. As witnessed by these demonstrations and because privacy guarantees must apply to the worst-case outliers (not only the average), any strategy for protecting data privacy should prudently assume that attackers have unfettered access to external data sources as well as internal model parameters.

---

[1] http://www.imdb.com/

UGC data has been used in many research areas ranging from psychology (e.g., predicting personality [5,6]) to data analytics (e.g., predicting stock market [7]). The privacy issues become more and more critical, especially when sensitive information (e.g., age, gender) can be derived from UGC data [8]. Thus, the main goal of this paper is to present a self-adaptive approach for privacy concern detection, which automatically detects the privacy need of individuals based on personality information extracted from their UGC data. In this way, we provide trade-off of sufficient privacy protection and data utility. The **main contributions** of this paper include:

- Introducing a neural network model that can learn and automatically predict the privacy-concern degree of individuals based on their personalities.
- Evaluating the effectiveness of personality based privacy-guarantee through extensive experimental studies on a real UGC dataset.
- Solving an imbalanced data distribution issue in privacy-concern detection raised by Vu et al. [9] using an over-sampling approach.

The remainder of this paper is organized as follows: In Section 2, we present related work, and introduce Differential Privacy [10] and the Five Factor Model [11]. Our proposed methodology is presented in Section 3. The experimental methodology is explained in Section 4. Experimental result analysis and discussion are in Section 5. Section 6 concludes the paper and presents future work.

## 2   Related Work

Anonymization [12] and sanitization [13] have been widely used in privacy protection. Differential privacy [10] later emerged as the key privacy-guarantee mechanism by providing rigorous, statistical guarantees against any inference from an adversary. Based on differential privacy, some privacy-oriented frameworks arose including PINQ [14] and GUPT [15]. Moreover, they use a unified amount of noise for privacy protection. To overcome the disadvantage of injecting uniform noise in differential privacy, recent works [16,17] have proposed personalized differential privacy methods, in which they apply different amounts of noise on different users. However, the limitation is that they decided privacy budget by either random sampling or query involvements regardless of the provenance individual's actual privacy concern. Particularly, Jorgensen et al. [17] use random sampling for personalized DF and Hamid et al. [16] consider what records the given query involves to decreases the corresponding privacy budgets. Thus, we propose a personality-based differential privacy approach in a self-adaptive way to calculate privacy concern for reasonable privacy protection and data utility.

### 2.1   Differential Privacy Preliminaries

Differential privacy (DP) [10] has established itself as a strong standard for privacy preservation. It provides privacy guarantees for algorithms analyzing databases, which in our case is a machine learning algorithm processing a training dataset and histogram-based data analysis. The key idea behind differential privacy is to obfuscate an individual's properties, but not the whole group's properties in a given database. So the probability for any individual in the database needs to have a property that should barely differ from the base rate (i.e., the chance to guess whether an individual is involved in one study or not). When an attacker analyzes the database, he/she cannot reliably learn anything new about any individual in the database, no matter how much additional information he/she has. The following is a formal definition of ($\epsilon$-$\delta$) differential privacy. We assume a database $D$ consisting of $n$ vectors of $m$-components over some set $\mathcal{F}$ represented as a $m \times n$ matrix over $\mathcal{F}$.

**Definition 1 (Distance Between Databases).** Define

$$\mathrm{dist}(D, D') := |\{i \in \{1, 2, \ldots, m\} \colon D_i \neq D'_i\}| \, \forall D, \, D' \in (\mathcal{F}^m)^n$$

as the number of entries in which the databases $D$ and $D'$ *differ*. Differential privacy is defined using pairs of adjacent databases in present work, which only differ by one record (i.e., $\mathrm{dist}(D, D') = 1$).

**Definition 2 (Probability Simplex).** Given a discrete set $B$, the probability simplex over $B$, denoted $\Delta(B)$ is defined to be:

$$\Delta(B) = \left\{ x \in \mathbb{R}^{|B|} : x_i \geq 0 \text{ for all } i \text{ and } \sum_{i=1}^{|B|} x_i = 1 \right\}$$

A randomized algorithm with domain $A$ and (discrete) range $B$ will be associated with a mapping from $A$ to the probability simplex over $B$, denoted as $\Delta(B)$.

**Definition 3 (Randomized Algorithm).** A randomized algorithm $\mathcal{M}$ with domain $A$ and discrete range $B$ is associated with a mapping $\mathcal{M} : A \to \Delta(B)$. On input $a \in A$, the algorithm $\mathcal{M}$ outputs $\mathcal{M}(a) = b$ with probability $(\mathcal{M}(a))_b$ for each $b \in B$.

**Definition 4 (($\epsilon$-$\delta$)-differential privacy).** Let $\mathcal{M}$ be a randomized algorithm processing $D$ and $\mathrm{Range}(\mathcal{M})$ its image. Now $\mathcal{M}$ is called ($\epsilon$-$\delta$)-differentially private if $\forall \mathcal{S} \subseteq \mathrm{Range}(\mathcal{M})$:

$$\forall D, D' \colon \mathrm{dist}(D, D') \leq 1 \Rightarrow \Pr\left[\mathcal{M}(D) \in \mathcal{S}\right] \leq e^\epsilon \cdot \Pr\left[\mathcal{M}(D') \in \mathcal{S}\right] + \delta$$

Intuitively, differential privacy controls the degree to which $D$ and $D'$ can be distinguished. When $\delta = 0$ then ($\epsilon$-$\delta$)-differential privacy is also called $\epsilon$-differential privacy. Smaller $\epsilon$ gives more privacy and lower utility. Then, given the result of a randomized algorithm $\mathcal{M}$, an attacker cannot learn any new property about data subjects with a significant probability.

**The Global Privacy Budget** PINQ [14] is an implementation of interactive differential privacy which ensures, at runtime, that queries adhere to a global privacy budget $\epsilon$. Its central principle is that multiple queries (e.g., with differential privacy $\epsilon_1$ and $\epsilon_2$ respectively) have an additive effect $\epsilon_1 + \epsilon_2$ on the overall differential privacy. In other words, for a dataset queried $q$ times, with each query having privacy parameter $\epsilon_i$, the total privacy budget of the dataset is given by $\epsilon_{total} = \sum_{i=1}^{q} \epsilon_i$. PINQ also tracks sensitivity of functions to track how much to deduct from the global privacy budget on each invocation of a primitive query. As mentioned in [16], the global privacy budget has limitations when applied to an interactive system: (1) data analysts using the system, may run out of privacy budget even before obtaining valuable results and (2) a global budget is not capable of handling a live database when new records are frequently added.

## 2.2  The Five Factor Model

Regarding personality prediction, the most influential Five Factor Model (*FFM*) has become a standard model in psychology over the last 50 years [11]. Here we re-introduce a summary of Vu et al. [9] regarding FFM. The five factors are defined as *neuroticism, openness to experience, conscientiousness, agreeableness*, and *extraversion*. Pennebaker et al., [5] identify many linguistic features associated with each of personality traits in *FFM*. (1) *Extroversion* (sEXT) tends to seek stimulation in the external world, the company of others, and to express positive emotions. (2) *Neurotics* (sNEU) people use more 1st person singular pronouns, more negative emotion words than positive emotion words. (3) *Agreeable* (sAGR) people express more positive and fewer negative emotions. Moreover, they use relatively fewer articles. (4) *Conscientious* (sCON) people avoid negations, negative emotion words and words reflecting discrepancies (e.g., should and would). (5) *Openness to experience* (sOPN) people prefer longer words and tentative expressions (e.g., perhaps and maybe), and reduce the usage of 1st person singular pronouns and present tense forms.

## 3  Methodology

As mentioned above, one limitation of differential privacy is the unified privacy budget on all individuals in the same dataset. To address this limitation, we propose a personality-based differential privacy algorithm. The proposed approach is motivated by the findings on the statistically verified correlation between personality and privacy concerns of individuals on Facebook [18]. In their work, they output the correlation values as *p-value* = {.003, .007, .010} for {cNEU, cEXT, cAGR} personality traits accordingly. Sumner et al., however, did not mention about the *p-value* of the correlation between privacy-concern and {cCON, cOPN} personal traits. Moreover, this personality-based privacy can be characterized as personalized-differential privacy that also satisfies $\epsilon$-differential privacy by the proof of Ebadi et al. [16]. However, Ebadi et al. did not have an automatic way of detecting personalized privacy-concern level, which we are addressing.

**Problem Definition**

Given a database $\mathcal{D}$ consists of $N$ user records $U = \{T, P\}$, where $T$ is a set of textual features and $P$ is a set of five personality trait scores in the Five Factor Model (FFM) [19]. Our target is predicting a real value $r$ that represents the user privacy-concern degree. The $r$ value, later on, will be used to decide the amount of noise and privacy-budget of the user to protect her/his data privacy more reasonably.

**A Baseline Linear Regression Model**

As a baseline linear regression model, we learn a linear function $y = w * \bar{x} + b$ where $\bar{x}$ is an input vector and $b$ is a bias value. Given a real outcome $\hat{y}$, we can calculate a loss function $\mathcal{L} = \frac{1}{2} \sum_{i=1}^{N}(y_i - \hat{y}_i)^2$, where $N$ is the number of samples and use it for the optimization process to find the best values of the weighting matrix $w$, i.e., $w^* = \arg\min_{w} \mathcal{L}(w)$. Since we have five personality scores, therefore, five linear regression models are learned to predict scores of the five personality traits (i.e., *neuroticism, openness to experience, conscientiousness, agreeableness, and extraversion*). Lastly, we directly adopt a scaled weighting vector $V = (5.0, 4.7, 4.3, 4.1, 1.0)$ from [18] to calculate $r = sigmoid(z)$, where $z = \sum_{i=1}^{5} v_i * y_j$, $V = (v_i \mid i = 1, 2 \ldots, 5)$ and $Y = (y_j \mid j = 1, 2 \ldots, 5)$ is the output vector when predicting the five personality traits.

**A Deep Neural Network Regression Model**

We hypothesize that the five personality traits are highly correlated, therefore, a jointly learning regression model will lead to a better prediction model compared to the five standalone linear regression models. Thus, differently from the baseline, we design a method, a multilayer perceptron neural network model (MLP) [20] to learn and predict $r$ value directly (see Figure 1).

Formally, a one-hidden-layer MLP is a function $f : R^K \to R^L$, where $K$ is the size of input vector $x$ and $L$ is the size of the output vector $f(x)$, such that, in matrix notation: $f(x) = G(b^{(2)} + W^{(2)}(s(b^{(1)} + W^{(1)}x)))$, with bias vectors $b^{(1)}, b^{(2)}$; weight matrices $W^{(1)}, W^{(2)}$ and activation functions $G$ and $s$. The vector $h(x) = \Phi(x) = s(b^{(1)} + W^{(1)}x)$ constitutes the hidden layer. $W^{(1)} \in R^{K \times K_h}$ is the weight matrix connecting the input vector $x$ to the hidden layer $h$. Each column $W^{(1)}_{\cdot i}$ represents the weights from the input units to the $i$-th hidden unit. The output vector is then obtained as: $o(x) = G(b^{(2)} + W^{(2)}h(x))$. To train an MLP, we learn all parameters of the model, and here we use Adam optimizer [21] with minibatches to control the learning rate. The set of parameters to learn is the set $\theta = \{W^{(2)}, b^{(2)}, W^{(1)}, b^{(1)}\}$. The gradients $\partial \ell / \partial \theta$ can be obtained through the backpropagation algorithm [22] (a special case of the chain-rule of derivation). Lastly, in the linear layer, we apply a similar approach to the baseline linear regression method to calculate the final prediction.

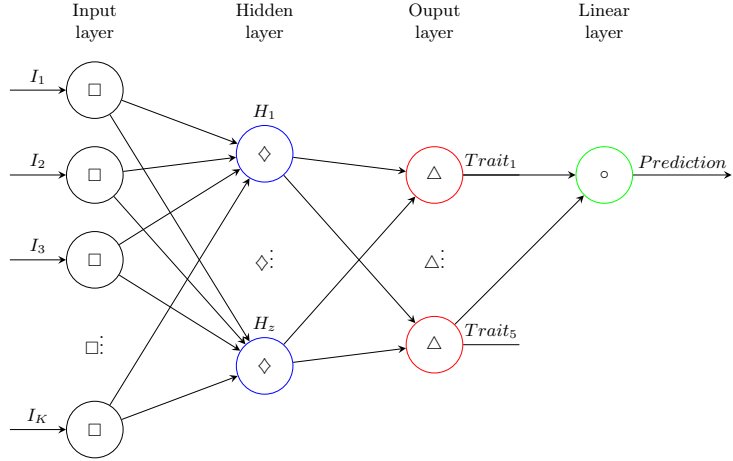

**Figure 1.** Network graph of a multilayer perceptron with $K$ input units, $L$ output units ($L$=5), and a linear layer.

**Implementation details**: we use Tensorflow [23] to implement our model. Model parameters are learned to minimize the cross-entropy loss with $L_2$ regularization. Table 1 shows optimal settings for the regression task and Figure 2 shows the learning curve of the MLP model on training data following 10-fold cross-validation schema (i.e., each fold, 10% is used for validation).

| Name | Value |
|---|---|
| Hidden layers | 80 |
| # epoch | 90 |
| Learning rate | $10^{-5}$ |
| $l_2$ | $10^{-5}$ |

**Table 1.** The optimal hyperparameter settings.

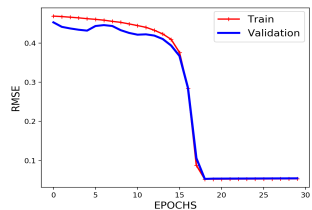

**Figure 2.** Learning curve of the MLP.

## 4    Experimental Methodology

**Dataset** We evaluate our methods on myPersonality data[2]. It contains personality scores and Facebook profile data, collected by Stillwell and Kosinski by means of a Facebook application that implements the FFM personality questionnaire in a 100-item long version of [24]. The application obtained the consent from its users to record their data and use it for the research purposes. They selected only the users for which they had both information about personality and social network structure. The status updates have been manually anonymized. The final dataset contains 9,917 Facebook statuses of 250 users in raw text, gold standard (self-assessed) personality labels, and several social network measures. Figure 3 shows the relations between different features in the data. As we can see a higher correlation between openness personal trait to others. Moreover, the data distribution of sOPN and sNEU in the dataset are more imbalanced than other personal traits.

---
[2] http://myPersonality.org

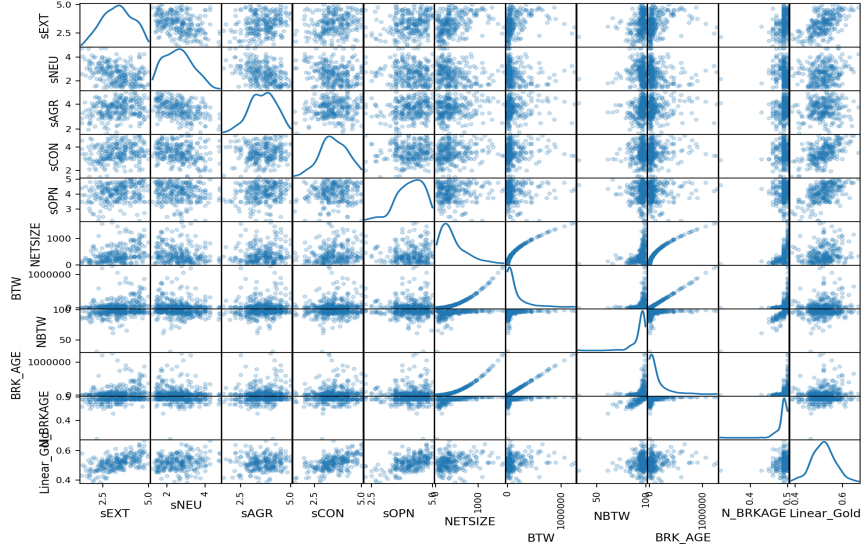

**Figure 3.** Correlation between different features in the dataset excluding Facebook status updates. The features here are network size (NETSIZE), betweenness (BTW), the number of betweenness (NBTW), density (DEN), brokerage (BRK_AGE), the number of brokerage (NBRKAGE), and *Linear_Gold* is the implicit gold regression values.

### 4.1 Gold Standard Values

Ideally, we would evaluate downstream performance compared to a ground truth. Unfortunately, a ground truth is difficult to characterize for the privacy-concern task since people would have answered "as high as possible" if someone simply asked them "how much privacy-guarantee do you want to have?". Our future work would be connecting computer science, crowdsourcing and psychology in order to collect gold standards on user privacy concern using psychological and behavior tests. Since this type of ground truth does not exist currently, we constructed the gold standards in our work as follows, using all available information about users to be an approximation of the ground truth.

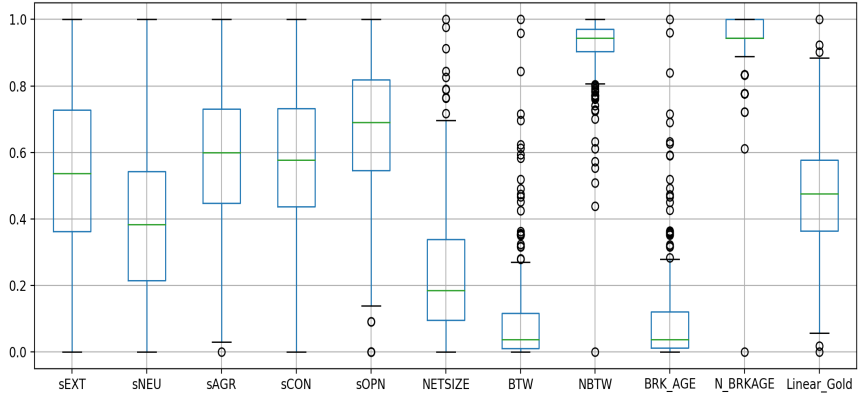

**Figure 4.** The distribution of numerical features using box plots. All numbers are normalized to stay between [0, 1] in order to have a good overview of the whole features.

**Gold Regression Values** The myPersonality dataset contains both personality trait labels and personality trait scores. Therefore, previous works in personality profiling can be catergorized as classification problems [25] or regression problems [26,27]. Since there is no gold privacy-concern labels/values in the dataset, we implicitly derive the values using a linear combination between personality trait scores and the weighting vector $V$ adopt from [18]. We denote $P = (r_i \mid i = \{1, 2 \ldots N\})$, where $N$ is the number of samples and $r$ is the real privacy concern degree of a certain user. Then, $r = \frac{1}{125} * \sum_{i=1}^{5} v_i * s_j$, where $V = (v_i \mid i = 1, 2 \ldots, 5) = (5.0, 4.7, 4.3, 4.1, 1.0)$ taken from [18], and 125 is a normalisation constant. $S = (s_j \mid j = 1, 2 \ldots, 5)$ is user personality trait scores. $P$ is also denoted as *Linear_Gold* in both Figure 3 and 4 (i.e., the last column).

**Gold Classification Labels** We constructs gold labels to evaluate downstream classification performance followed the work of Vu et al. [9]. The labels are high (HiPC), medium (MePC), and low privacy-concern (LoPC) level as firstly proposed in [16]. Given the ground truth of personality labels, i.e., **yes** and **no** labels of {NEU, OPN, CON, AGR, EXT} - the five personality traits. Based on the findings of [18] we know that privacy concerns of different personality traits are ordered as following {NEU, OPN, CON, AGR, EXT} from the highest privacy-concern to the lowest privacy-concern correspondingly. Thus, we derive the privacy concern levels as in Table 2. Eventually, our ground truth set consists of 29 users in HiPC, 212 users in MePC, and 9 users in LoPC.

| Privacy concern label | cNEU | cOPN | cCON | cAGR | cEXT |
|---|---|---|---|---|---|
| HiPC | **yes** | **yes** | any | no | no |
| LoPC | no | no | any | **yes** | **yes** |

**Table 2.** Deriving self ground-truth labels for classification problem. It is note that MePC are the rest, i.e., {¬HiPC, ¬LoPC}.

### 4.2  Feature Extraction

Feature extraction is a process of extracting valuable and significant information from the raw data to represent the data. Since collecting personally sensitive data is cautious and challenging, the myPersonality dataset is small [28], so we have to incorporate with pre-trained embedding models (e.g., Word2Vec[3]) to better represent the data. Table 3 lists all extracted features in this work inspired by Vu et al. [29]:

| Feature | # Features | Information |
|---|---|---|
| Lexical Features | 7111 | N-grams features, i.e., [1,2,3,4,5]-grams. |
| Topic Features | 200 & 50 | For LSI and LDA features respectively. |
| Semantic features | 300 | Word2Vec model trained on Google News. |
| Total | 7661 | |

**Table 3.** Total number of extracted features in this work

### 4.3  Evaluation Results

We design two different types of experiments to evaluate our methodology. Firstly, we evaluate how well we can detect privacy-concern regarding both classification problems and regression problems. Secondly, we analyze the effect of personality-based privacy controller to see if our self-adaptive approach can better balance data utility and privacy preservation. Abbreviations used in this section are listed in Table 4.

**Privacy-Concern Detection** Using the above ground truth data, similar to [9], we build two different privacy-classifiers with Naive Bayes and Support Vector Machine (SVM) algorithms for the classification task. Table 5 shows the performance of privacy-concern detection in comparison with the work of Vu et al. [9]. In their work, the authors showed that due to the imbalance of class distribution, Naive Bayes (NB) does not perform well. In this work, we solve the imbalanced data distribution using [30]. Table 6 shows the data distribution before and after the over-sampling process. Thus, NB and SVM both perform much better than the majority accuracy.

---
[3] https://code.google.com/archive/p/word2vec/

| # | Abbreviation | Description |
|---|---|---|
| 1 | RMSE | Root Mean Square Error, $\text{RMSE}(y, \hat{y}) = \sqrt{\frac{\sum_{t=1}^{n} (\hat{y}_t - y_t)^2}{n}}$ |
| 2 | EVS | Explained Variance Score, $\text{EVS}(y, \hat{y}) = 1 - \frac{Var\{y - \hat{y}\}}{Var\{y\}}$ |
| 3 | OOBudget | Out of budget records saying number (ratio) of records that get out of budget at a certain iteration. |
| 4 | MLP-OOBudget | OOBudget of the multi-layer perceptrons algorithm |
| 5 | SVR-OOBudget | OOBudget of the Support Vector Regression algorithm |
| 6 | LR-OOBudget | OOBudget of the Linear Regression algorithm |

**Table 4.** List of abbreviation used in this section

Table 7 shows evaluation results of the regression-based problem with three different algorithms including: LR (Linear regression [31]), SVR (SVM regression [31]), and MLP (our proposed multi-layer perception neural network model with a linear layer). The data was divided to 80% for training and 20% for testing. Clearly, LR performs well on the training process, however, it was over-fitted the data (i.e., RMSE = 0, variance score = 1) and could not generalize well to predict testing data. In contrast, MLP performs reasonably well on the training data and achieves the best performance on the test data. It is worth to mention that personality trait scores were used to derive the *Linear_Gold* values (i.e., $P$), therefore, they are not included in the feature extraction process. This approach is also closer to the real scenario where we can easily collect UGC but not their personality scores.

| Paper | Vu et al. [9] | | | This work | | |
|---|---|---|---|---|---|---|
| Algorithm | Majority | Naive Bayes | SVM | Majority | Naive Bayes | SVM |
| Accuracy | 0.78 | 0.57 | 0.80 | 0.848 | **0.97** | 0.967 |

**Table 5.** Privacy concern detection performance in comparison with majority accuracy. The evaluation accuracy of this work is the average accuracy of 5-fold cross-validation.

| | Before over-sampling | | | After over-sampling | | |
|---|---|---|---|---|---|---|
| Labels | LoPC | MePC | HiPC | LoPC | MePC | HiPC |
| # of samples | 9 | 212 | 29 | 212 | 212 | 212 |

**Table 6.** Label distribution before and after over-sampling using SMOTE [30] to solve the imbalanced data distribution issue.

**Privacy-budget Controller** We design a learning task using SVM with the privacy-budget controller to see how it affects the classification performance. A 10-fold SVM classification is designed to interactively request valid user records until it receives no records. Thus, this test is similar to a real scenario where an analyst requests to the system and retrieves information. Figure 5 shows four

Self-adaptive Privacy Concern Detection for User-generated Content     11

|          | Algorithm | RMSE  | EVS     |
|----------|-----------|-------|---------|
|          | LR        | 0     | 1       |
| Training | SVR       | 4.460 | -8834   |
|          | MLP       | 0.058 | -0.472  |
|          | LR        | 0.064 | -0.602  |
| Testing  | SVR       | 6.277 | -15252  |
|          | MLP       | **0.052** | **-0.305** |

**Table 7.** Evaluation results of regression-based privacy concern detection. Note that the higher values of EVS are, the better (with the best possible score of 1.0).

different privacy budget controls including (a) global privacy budget, (b) random privacy budget, (c) linear regression based privacy budget, and (d) MLP privacy based budget.

Based on the experimental results, we have the following observations:

(1) Data utility of the global privacy budget quickly drops to 0 due to all records run out of privacy budget at the same time.
(2) Except of global privacy budget, the other three personalized-privacy budgets have better trade-off of privacy and data utility since they can avoid a situation where all records run out of privacy-budget at the same time.
(3) The random privacy-budget achieves better results in terms of data utility, however, it does not take into account the user privacy-concern level.
(4) MLP privacy-budget certainly shows a better way of controlling privacy and data utility. The privacy-budget gradually increases which allows data analysts, i.e., a classification algorithm in our experiment, to receive enough data records to maintain good classification results.

To the regression problem, Figure 6 shows evaluation results on 50 instances of the testing data. Our proposed MLP-budget controller clearly works better than others in terms of the following criteria:

(1) Regarding performance: we consider the Gold-RMSE (see Figure 6-(a)) is the standard. Comparing to the Gold-RMSE, MLP-RMSE and SVR-RMSE have the same trend. However, the mean distance of MLP-RMSE to the Gold-RMSE is 9.487, which is smaller than that of SVR-RMSE, i.e., 11.113 (see Table 8).
(2) Regarding OOBudget: LR-OOBudget is much similar to Gold-OOBudget comparing to SVR-OOBudget and MLP-OOBudget. However, its LR-RMSE was the worst.

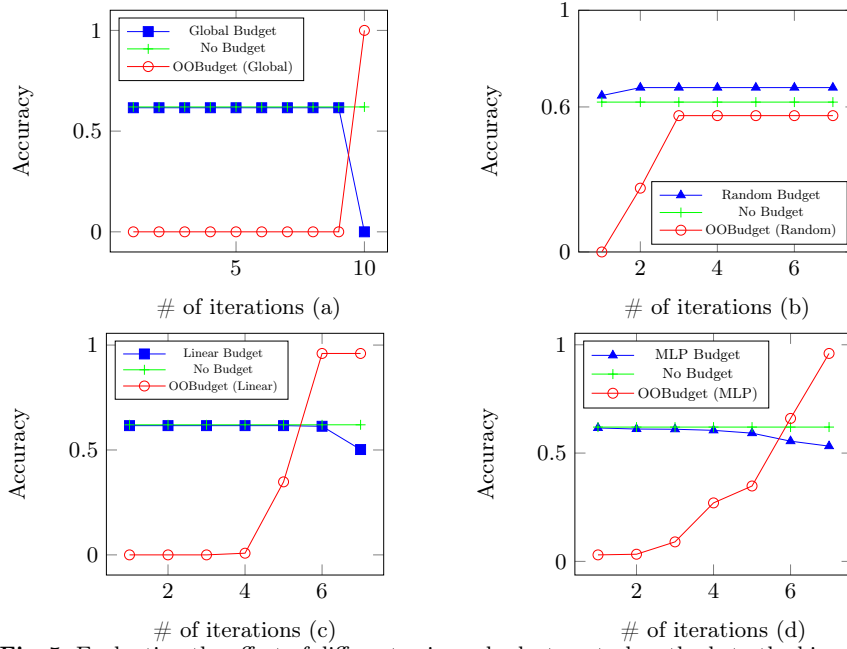

**Fig. 5.** Evaluating the effect of different privacy-budget control methods to the binary classification performance of the cEXT class. OOBudget is the ratio of out-of-budget user records.

| Algorithm & Iteration | 23 | 24 | 25 | 26 | 27 | 28 | 29 | 30 | Distance (RMSE) |
|---|---|---|---|---|---|---|---|---|---|
| Gold | 1 | 3 | 4 | 14 | 21 | 28 | 35 | 42 | 0 |
| LR | 0 | 1 | 4 | 7 | 13 | 26 | 38 | 45 | 4.183 |
| SVR | 0 | 0 | 0 | 0 | 1 | 39 | 49 | 49 | 9.487 |
| MLP | 0 | 0 | 0 | 0 | 4 | 26 | 48 | 48 | 11.113 |

**Table 8.** Distance to the gold regression in comparison to LR, SVR, and MLP. The iteration running from 1 to 30, but we only show from the iteration 23rd where we can observe OOBudget.

## 5   Conclusions and Future Work

This paper presents a self-adaptive differential privacy preserving approach for data analysis. To address the limitation of unified privacy budget in differential privacy, we calculate privacy budget based on personality knowledge. According to our experiments, personality-based privacy budget shows a more practical way of controlling privacy and data utility. Moreover, our proposed approach (i.e., MLP privacy budget) shows the best trade-off of data utility and privacy control. Our approach is applicable to real scenario where we only have user-generated content information. As mentioned before, there is no gold standard values (or labels) from users regarding their privacy-concern. Therefore, one of our future work directions is applying crowdsourcing technologies in user privacy concern detection to contribute the construction of a ground truth framework.

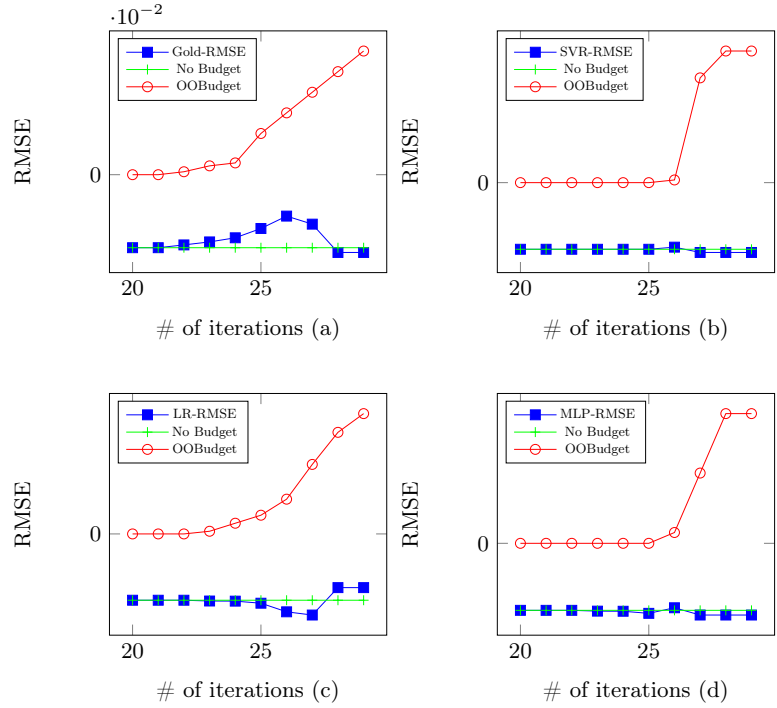

**Fig. 6.** Evaluating the effect of different privacy-budget control methods to the regression problem performance of the testing data. OOBudget is the ratio of out-of-budget user records.